\documentstyle[aps,twocolumn,prb]{revtex}
\begin{document}

\twocolumn[\hsize\textwidth\columnwidth\hsize\csname@twocolumnfalse\endcsname

\title{Dynamical Properties of Two Coupled Hubbard Chains \\ at Half--filling}

\author{ H. Endres$^a$, R.M. Noack$^a$, W. Hanke$^a$, D. Poilblanc$^b$,
and D.J. Scalapino$^c$ }

\address{ a) Institut f\"ur Theoretische Physik, Universit\"at
W\"urzburg, \\ Am Hubland,  97074 W\"urzburg, Germany}

\address{ b) Universit\'e Paul Sabatier, Laboratoire de Physique
Quantique,  \\ 118 route de Narbonne, 31062 Toulouse, France }

\address{ c) Department of Physics, University of California, Santa
Barbara, CA 93106}

\maketitle
\begin{abstract}
Using grand canonical Quantum Monte Carlo (QMC) simulations combined
with Maximum
Entropy analytic continuation, as well as analytical methods, we
examine the one-- and two--particle
dynamical properties of the Hubbard model on two coupled chains
at half--filling.
The one--particle spectral weight function,
$A({\bf k},\omega)$,  undergoes a qualitative change with interchain
hopping $t_\perp$
associated with a transition from a four--band insulator to a two--band
insulator.
A simple analytical model based on the propagation of exact rung
singlet states gives a good
description of the features at large $t_\perp$.
For smaller $t_\perp$, $A({\bf k}, \omega)$ is similar to that of the
one--dimensional model, with a coherent band of width the effective
antiferromagnetic exchange $J$ reasonably well--described by
renormalized spin--wave theory.
The coherent band rides on a broad background of width several times the
parallel hopping integral $t$, an incoherent structure similar to
that found in calculations on both the one-- and two--dimensional
models.
We also present QMC results for the two--particle spin and charge
excitation spectra,
and relate their behavior to the rung singlet picture for large
$t_\perp$ and
to the results of spin--wave theory for small $t_\perp$.
\end{abstract}
]


\section{\label{INTRO} INTRODUCTION}

The compounds SrCu$_2$O$_3$ \cite{takano} and (VO)$_2$P$_2$O$_7$
\cite{johnston}
consist of arrays of weakly interacting two--chain metal--oxide ladders.
In SrCu$_2$O$_3$, CuO$_2$ ladders are coupled by a weak, frustrated,
ferromagnetic coupling\cite{rice}, and in (VO)$_2$P$_2$O$_7$,
VO$_4$ ladders are well separated in the structure of the material.
These materials have a half--filled conduction band,
corresponding to one conduction electron per metal--oxide unit and are
insulators with short--range antiferromagnetic correlations and a spin
gap.
Very recently Z.\ Hiroi and M.\ Takano\cite{hit95} succeeded in doping
the two--leg ladder compound La$_{1-x}$Sr$_x$CuO$_{2.5}$.
They observe a fall in resistivity with doping, possibly to a metallic
state, but no sign of superconductivity. The spin--gap found in
half--filled LaCuO$_{2.5}$ persists, but decreases with doping.
In this publication we will address the half--filled situation only.
The undoped materials have been described by
a Heisenberg model on two coupled chains \cite{rice,dagotto,barnes1,whiteprl},
which explains a number of experimentally observed magnetic
properties of the materials, including the size of the spin gap, and
the spectrum of the spin triplet excitations.
The Hubbard model on two coupled chains, which at half--filling and in
the limit of strong on--site Coulomb repulsion can be
mapped onto the Heisenberg model, gives an itinerant electron picture
of these systems \cite{nws94} which includes charge as well as spin
fluctuations.
While the ground state and the low--energy properties of the
two--chain Heisenberg
and half--filled Hubbard model are becoming well understood, less is
known about the finite frequency one-- and two--particle response.
The dynamic spin response of the Heisenberg model was studied in
Ref.\ 5 
and the single particle and spin response was examined for a
$t$--$J$ model with two holes in Ref.\ 8 
using analytic continuation of Lanczos
exact diagonalization calculations on small clusters.
The dynamic properties are important because they can be measured with
a variety of experimental techniques.
For example, the one--particle spectral weight can be probed by
photoemission and inverse photoemission, the dynamic charge
correlation function by optical response measurements, and the dynamic
spin structure factor by inelastic neutron scattering.

Here we examine the dynamic one-- and two--particle correlation
functions of the two--chain Hubbard model at half--filling using
Quantum Monte Carlo simulations analytically continued to real
frequencies using a Maximum Entropy technique\cite{maxent}.
We study the dynamic response for large, intermediate, and small
values of the perpendicular hopping $t_\perp$.
This is interesting for two reasons:
first, when the electron--electron interaction is turned off ($U=0$)
in the two--chain system, there is a metal--insulator transition from
a two--band metal to a band insulator due to the separation of the bonding
and antibonding bands with increasing $t_\perp$.
For the interacting ($U \ne 0$) case, it is generally believed that the
two--chain system is always insulating (see, for example, Ref. 11 
for a discussion within a weak--coupling RG picture).
The QMC simulations do find an insulator for all $t_\perp$, but also
reveal a crossover from four--band
insulating behavior for small $t_\perp$ to two--band behavior for
\mbox{$t_\perp/t$ {\raise3pt\hbox{$>$}\llap{\lower3pt\hbox{$\sim $}}} $2$}.
A similar crossover occurs for weak coupling
($U$ {\raise3pt\hbox{$<$}\llap{\lower3pt\hbox{$\sim $}}} $t$)
within
antiferromagnetic Hartree--Fock (AFHF) theory, as we shall discuss in
section \ref{AFHFSEC}.
One interesting and important result of this paper is that this
four--band to two--band crossover scenario survives in the
intermediate to
large $U$ regime in the system, despite the fact that the
many--body physics for
$t_\perp$ {\raise3pt\hbox{$>$}\llap{\lower3pt\hbox{$\sim $}}} $t$
is not even qualitatively
reproduced in the AFHF approximation.
No such crossover is present as a function of $J_\perp/J$ in the
Heisenberg model, because the
mapping to the Heisenberg model breaks down when
$J_\perp \sim 4 t_\perp^2/U$ is of the order of $t_\perp$.
In other words, the effect of the electronic band structure is not
present in the Heisenberg model.
As we shall see, this crossover can also be described in terms of the
crossover from a localized rung picture, in which localized
excitations form a Bloch state along the chains
for large $t_\perp$, to a picture with longer range
antiferromagnetic correlations with many features that can be
qualitatively described by
AFHF 
spin--wave theory for small
$t_\perp$.

Secondly, in the small to intermediate $t_\perp$ regime, this
system displays features of the single--particle spectral weight
$A({\bf k},\omega)$ seen
in recent numerical treatments of both the 1D\cite{preuss1d,newpreuss1d} and
2D\cite{preuss2d,dagotto2d} Hubbard model.
We find two features also generic to the 1D and 2D systems:
a dispersive band whose width is of the order
of the effective exchange interaction $J\sim 4t^2/U$
and an incoherent background of width several times the parallel
hopping integral $t$.
For the one and two chain systems, but not the 2D system, the
dispersive band is well described by renormalized spin--wave--theory.

We consider the  single band Hubbard model on two coupled chains of
length $L$:
\begin{eqnarray}
H&=&-t\sum_{i,\lambda\sigma} (c^\dagger_{i,\lambda\sigma}
               c^{\phantom{\dagger}}_{i+1,\lambda\sigma} + h.c.)
\nonumber\\
 && -t_\perp \sum_{i,\sigma} (c^\dagger_{i,1\sigma}
               c^{\phantom{\dagger}}_{i,2\sigma} + h.c.)
    +U\sum_{i\lambda} n_{i,\lambda\uparrow}
 n_{i,\lambda\downarrow}.
\label{hamiltonian}
\end{eqnarray}
Here $c^\dagger_{i,\lambda\sigma}$ and
$c^{\phantom{\dagger}}_{i,\lambda\sigma}$ create and annihilate electrons
on rung $i$ and chain $\lambda$ with spin
$\sigma$, respectively,
the hopping integral parallel to the chains is
$t$, the hopping between the chains $t_\perp$, and $U$ is an on-site
Coulomb interaction.
We use periodic boundary conditions parallel to and
open boundary conditions perpendicular to the chains.

The non--interacting, $U=0$, Hamiltonian can be diagonalized by
writing the hopping term in terms of bonding and antibonding states on
a rung and Fourier--transforming parallel to the chains.
The energy is then given by
\begin{equation}
\varepsilon_{\bf k} = -(2t \cos k + t_\perp \cos k_\perp)
\label{eqepsilon}
\end{equation}
with ${\bf k}=(k,k_\perp)$,
where $k_\perp=0$ and $k_\perp=\pi$ corresponds to the energy of the
bonding and antibonding band, respectively, and $k$ is the momentum
along the chains.
Both bands will be occupied when $t_\perp < t_{\perp c}$,
whereas when $t_\perp > t_{\perp c}$, only the bonding band will be
occupied.
Here $t_{\perp c}/t \equiv 1 - \cos \pi \langle n \rangle$ and
$\langle n \rangle \equiv \langle \sum_\sigma c^\dagger_{i,\lambda,\sigma}
c_{i,\lambda,\sigma}\rangle$.
For half--filling $t_{\perp c}/t=2$, and the system is
a two--band metal for $t_\perp/t < 2$, and a band insulator
with a completely occupied bonding band for $t_\perp/t > 2$.

\section{\label{SINGLE} SINGLE--PARTICLE SPECTRAL WEIGHT}

\subsection{ Quantum Monte Carlo Results}

In order to calculate the dynamical properties of the
above model we employ the grand--canonical Quantum
Monte Carlo (QMC) algorithm to determine the
expectation values of the correlation functions in imaginary time.
At half--filling, the simulations are not limited by the fermion sign problem
and accurate results can thus be achieved at low temperatures and large
system sizes.
We analytically continue the imaginary--time data to real frequencies
using a Maximum Entropy method\cite{maxent}. For example, the spectral weight
function given by
\begin{eqnarray}
A({\bf k},\omega) &=& {1\over Z}
\sum_{l,l^\prime} e^{-\beta E_l}(1+e^{-\beta\omega})
\vert \langle l\vert c_{{\bf k},\uparrow}\vert l^\prime \rangle \vert^2
\nonumber\\
&& \phantom{{1\over Z}\sum_{l,l^\prime}}
\cdot\delta(\omega-(E_{l^\prime}-E_l)),
\label{Akomegadef}
\end{eqnarray}
where $Z$ is the partition function, $\vert l>$ and $\vert l^\prime>$ are
the exact many-body eigenstates with the corresponding energies $E_l$ and
$E_{l^\prime}$ and $c_{{\bf k},\sigma}=\sum_{i,\lambda}
c_{i,\lambda\sigma} e^{i{\bf k}\cdot{\bf R}_{i,\lambda}}/\sqrt{2L}$,
can be calculated by inverting the spectral theorem
\begin{equation}
G({\bf k},\tau)\equiv \langle c^{\phantom{\dagger}}_{{\bf k},\uparrow}(\tau)
                c^\dagger_{{\bf k},\uparrow}(0) \rangle
               = \int_{-\infty}^{\infty}
                {e^{-\tau\omega}\over 1+e^{-\beta\omega}}
                A({\bf k},\omega) d\omega
\label{eqGtaudef}
\end{equation}
with the Maximum Entropy algorithm using the raw QMC data
$G({\bf k},\tau)$.
To achieve high resolution, it is important to use a likelihood
function which takes the error--covariance matrix of the QMC data and
its statistical inaccuracy consistently into account\cite{preussmaxent}.
The results presented here are based on QMC data with good statistics,
i.e. averages over $10^5$ updates of all the Hubbard--Stratonovich
variables result in $G({\bf k}, \tau)$'s with absolute errors less
than or of the order of $5 \times 10^{-4}$.
Correlations of the data in imaginary time were taken into account by
making use of the covariance matrix in the Maximum Entropy procedure
\cite{preussmaxent}.
As suggested in previous work by White\cite{whi91}, various moments of
the spectral weight were also incorporated in extracting
$A({\bf k},\omega)$.

\begin{figure}
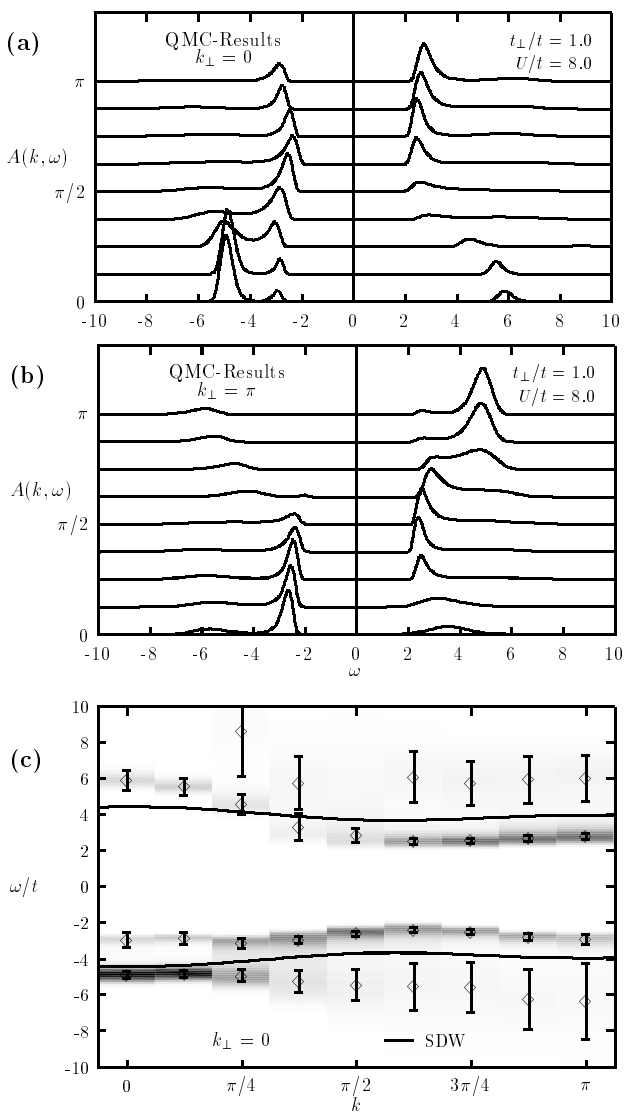

\caption{
        The single particle spectral weight $A({\bf k},\omega)$ as in
        Fig.\ 1, 
        with the same parameters except with an isotropic hopping,
        $t_\perp/t=1.0$.
        In (c), the solid line is the SDW result
        calculated within AFHF.
}
\label{figqmc1.0}
\end{figure}

We calculate the spectral function $A({\bf k},\omega)$ at half--filling
for a $2\times 16$-lattice with $U/t=8$ for different values of
$t_\perp$ at an inverse temperature of $\beta=10/t$.
Our results are given in Fig.\ \ref{figqmc2.0} for
$t_\perp/t=2.0$,
in Fig.\ \ref{figqmc1.0} for $t_\perp/t=1.0$,
and in Fig.\ \ref{figqmc0.5} for $t_\perp/t=0.5$.
Note that $t_\perp/t=1.0$ corresponds to the
isotropic case, $J_\perp\approx J$, relevant to the SrCu$_2$O$_3$,
(VO)$_2$P$_2$O$_7$, and  LaCuO$_{2.5}$ compounds.
Parts (a) and (b) in Fig.\ \ref{figqmc2.0}--\ref{figqmc0.5}
are three-dimensional plots of
$A({\bf k},\omega)$ versus $\omega$ for $k$--values in the 1D
Brillouin zone, whereas part (c) in all three figures summarize these
results in the usual ``band structure'' $\omega$ versus $k$ plot.
The shaded areas represent regions with significant spectral weight,
with the darker shading representing more weight.
Due to particle--hole symmetry,
$A({\bf k}=(k,\pi),\omega)=A({\bf k}=(\pi-k,0),-\omega)$ where
${\bf k}=(k,k_\perp)$.
In other words, one reflects $k$ about $k=\pi/2$ and $\omega$ about 0 to
get $A({\bf k},\omega)$ for $k_\perp=\pi$ from $A({\bf k},\omega)$ for
$k_\perp=0$.
This symmetry can be seen by comparing parts (a) and (b) of Fig.\
\ref{figqmc2.0}--\ref{figqmc0.5}.
Since this symmetry is not enforced by the Maximum Entropy
procedure, it is only present approximately, and the extent to which
the reflected spectra at $k_\perp=0$ do not match those at
$k_\perp=\pi$ gives an indication of the accuracy of the analytic
continuation procedure.
In the density plots in parts (c), we show only the $k_\perp=0$ components.
For $t_\perp/t=2.0$, there is a heavily weighted coherent band of
width $\sim 3t$ in the photoemission ($\omega < 0$) part of the spectrum.
In the inverse photoemission spectrum ($\omega > 0$) there is very
little total spectral weight. 
There are therefore two bands with significant spectral weight, one
in the photoemission spectrum for $k_\perp=0$ and one in the inverse
photoemission spectrum for $k_\perp=\pi$.
Each band is about $2t$ away from the Fermi surface, leading to a gap
of approximately $4t$.
Therefore, for large $t_\perp$, the system is a two-band insulator.

In contrast, for $t_\perp/t=1.0$ and $t_\perp/t=0.5$, $A({\bf k},\omega)$
has substantial spectral weight in four bands, one in the
photoemission spectrum ($\omega<0$) for $k_\perp=0$ and
$k_\perp =\pi$ and one in the inverse photoemission spectrum
($\omega >0$) for both values of $k_\perp$.
In this regime, the system is thus a four--band insulator.
For $t_\perp/t=0.5$ (Fig.\ \ref{figqmc0.5}) the spectral weight
is present for both values of $k_\perp$ concentrated between $k=0$
and $k=\pi/2$ in the photoemission part and between $k=\pi/2$
and $k=\pi$ in the inverse photoemission part of the spectrum.
For the isotropic case ($t_\perp/t=1.0$, Fig.\ \ref{figqmc1.0}), there
is some shift of spectral weight to the photoemission part for
$k_\perp=0$ and to the
inverse photoemission part for $k_\perp=\pi$.
The maxima of the photoemission band in the $k_\perp=0$ part
occurs at $k^*=\pi$ for $t_\perp/t=2.0$, at $k^*\approx 0.7\pi$ for
$t_\perp/t=1.0$, and at $k^*=\pi/2$ for $t_\perp/t=0.5$.
Therefore we would
expect to see a maxima at $k^*\approx 0.7\pi$ in a photoemission
experiment on the isotropic ladder compounds, an experiment which has,
to our knowledge, not yet been done.

It is also important to note that the QMC $A({\bf k},\omega)$ results for
$t_\perp/t=0.5$ as well as for the isotropic case, $t_\perp/t=1.0$,
display two general features which have recently been observed in both the
1D\cite{newpreuss1d} and 2D\cite{preuss2d} cases using
the improved Maximum Entropy techniques described above\cite{preussmaxent}.
One feature is that $A({\bf k},\omega)$ contains a rather
dispersionless ``incoherent background'' extending over several $t$
($\sim 6t$ for $U/t=8$ in 2D) in both the electronically occupied
($\omega < 0$) and unoccupied  ($\omega > 0$) parts of the spectrum.
The crucial structure, not resolved in previous 1D and 2D
QMC simulations, is a dispersive structure at low energies with a small
width of the order of the exchange coupling $J=4t^2/U$.
It is this latter coherent ``band'' which defines the gap $\Delta$ and
which, for larger $U$ ($U/t\sim 10$), is well--separated from the
higher--energy background.
As in the 1D and 2D cases, the splitting in the
low--energy band and the higher--energy background is
especially pronounced near $k=0$ and $k=\pi$ due to a relative weight
shift from negative to positive energies as $k$ moves through $k^*$.

\subsection{\label{AFHFSEC}Spin--Wave Theory}

\begin{figure}
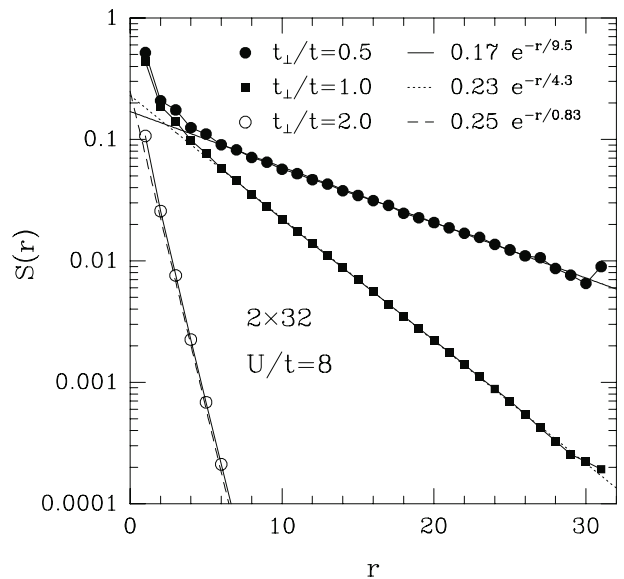

\caption{
        The spin-spin correlation function $S(r)$
	for $t_\perp=2t$, $t_\perp=t$ and $t_\perp=0.5t$
	calculated on a $2 \times 32$ lattice with $U/t=8$ and open boundary
	conditions using the DMRG.
	The solid, dotted and dashed lines are asymptotic fits to the data
	using the simple exponential form $A \exp(-r/\xi)$.
}
\label{figDMRG}
\end{figure}

At half--filling, the two chain Hubbard system at $T=0$ is a spin
liquid with a spin gap and a finite spin--spin correlation length.
The system is therefore less ordered than either the 1D system, which
is at the critical point and has power law decay of the spin--spin
correlation function and no spin gap, or the 2D system, which has
long--range spin order and gapless spin--wave--like excitations.
However, at large $U/t$ the scale of the spin gap for the two chain
system is set by $J_\perp \sim 4t_\perp^2/U$, and the spin--spin
correlation length $\xi$ grows longer as the spin gap gets smaller.
The range of the spin ordering can therefore be tuned by varying
$t_\perp/t$.
This is illustrated by Fig.\ \ref{figDMRG}, in which we plot the
spin--spin correlation function,
$S(r)=(-1)^r \langle M_{0,\lambda}^z M_{r,\lambda}^z \rangle$ on a
semilog scale for $t_\perp/t=2.0$, $t_\perp/t=1.0$ and $t_\perp/t=0.5$.
Here $M_{r,\lambda}^z=(n_{r,\lambda,\uparrow} - n_{r,\lambda,\downarrow})$
is twice the $z$ component of the on--site spin.
These calculations were done using the Density Matrix Renormalization
Group method\cite{nws94} (DMRG) on a system with open boundaries at zero
temperature.
For $t_\perp/t=0.5$, the data was averaged over a number of
different pairs of points for a given $r$ in order to reduce
oscillations due to the open boundaries.
For distances greater than a few lattice spacings, the correlation
functions are very well fit by a pure exponential form $A\exp(-r/\xi)$,
from which the spin--spin correlation length $\xi$ can be determined.
For $t_\perp/t=2.0$, $\xi=0.83$, for $t_\perp/t=1.0$, $\xi=4.3$,
and for $t_\perp/t=0.5$, $\xi=9.5$ lattice spacings.
Therefore, the spins are almost completely uncorrelated along the chains
for $t_\perp/t=2.0$, but for $t_\perp/t=1.0$ and for $t_\perp/t=0.5$
are correlated on length
scales of the order of the size of the $2\times 16$ lattice studied here,
although the long--range behavior is that of a gapped spin liquid in
all three cases.

It is therefore reasonable to calculate the single particle
spectral weight
$A({\bf k},\omega)$ in two simple ways:
first, when the spin gap is small and $\xi \approx L$, and we consider
wavelengths smaller than $\xi$ and frequencies larger than the spin
gap, it is useful to explore the consequences of a simple SDW
approximation based on an ordered state,
i.e. AFHF 
theory. Secondly,
one can solve for the exact eigenstates on a rung and then
consider states composed of a product of exact rung singlets only and a
delocalized one--hole state in a background of rung singlets.
(We call this the Local Rung Approximation, LRA.)
The LRA starts with a state with uncorrelated rungs and then treats
the interactions between rungs perturbatively, and is thus a good
starting point when $\xi$ is small.
As we shall see, the LRA gives a single particle spectral weight
distribution that agrees well with QMC in the large $t_\perp$ regime,
while AFHF qualitatively describes most features of the
spectral weight distribution for small $t_\perp$.

\begin{figure}
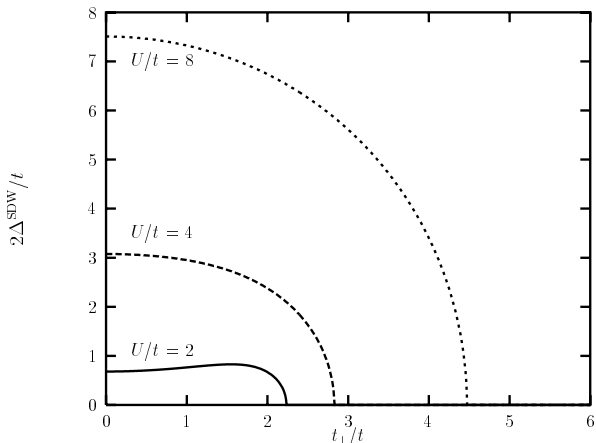

\caption{
	The SDW gap calculated within the AFHF approximation at
	$T=0$, as a function of $t_\perp/t$ on a $2 \times 100$ grid
	for $U/t=2,4,8$.
}
\label{figAFHFgap}
\end{figure}

We first discuss the AFHF calculation in more detail.
The Hartree-Fock one--particle Hamiltonian can be written as
\begin{equation}
H_{HF}={\sum_{{\bf k},\sigma}}^\prime E_{\bf k}
   (b^{c\dagger}_{{\bf k},\sigma} b^{c\phantom{\dagger}}_{{\bf k},\sigma}
 - b^{v\dagger}_{{\bf k},\sigma} b^{v\phantom{\dagger}}_{{\bf k},\sigma}),
\end{equation}
where ${\sum_{{\bf k},\sigma}}^\prime$ means that the sum runs only over
the magnetic zone ($-\pi/2 < k\le \pi/2$, $k_\perp=0,\pi$).
The operators $b^c_{{\bf k},\sigma}$ and
$b^v_{{\bf k},\sigma}$ are given by the following transformation\cite{swz89}:
\begin{eqnarray}
b^c_{{\bf k},\sigma} & = & u_{\bf k} c_{{\bf k},\sigma} \mp
                     v_{\bf k} c_{{\bf k}+{\bf Q},\sigma},\\
b^v_{{\bf k},\sigma} & = & v_{\bf k} c_{{\bf k},\sigma} \pm
                     u_{\bf k} c_{{\bf k}+{\bf Q},\sigma},
\end{eqnarray}
where ${\bf Q}=(\pi,\pi)$, the upper (lower) sign corresponds to
$\sigma=\uparrow$ ($\sigma=\downarrow$), and
\begin{eqnarray}
u_{\bf k} & = & \sqrt{{1\over 2}(1+\varepsilon_{\bf k}/E_{\bf k})},\\
v_{\bf k} & = & \sqrt{{1\over 2}(1-\varepsilon_{\bf k}/E_{\bf k})},\\
E_{\bf k} & = & \sqrt{\Delta^2+{\varepsilon_{\bf k}}^2}.
\label{eqAFHFdisp}
\end{eqnarray}
Here $\varepsilon_{\bf k}$ is the free particle energy ($U=0$) defined in
Eq.\ (\ref{eqepsilon}).
The spin--density--wave gap, $\Delta$, is
self--consistently determined by the equation
\begin{equation}
{1\over U} = {1\over 2L}{\sum_{\bf k}}^\prime {1\over E_{\bf k}}.
\label{eqAFHFgap}
\end{equation}
The band structure of the AFHF Hamiltonian is given by
$\pm E_{\bf k}=\pm\sqrt{\Delta^2+{\varepsilon_{\bf k}}^2}$ in the
magnetic Brillouin zone.
The SDW gap $\Delta$, calculated by solving Eq.\ (\ref{eqAFHFgap})
self--consistently, is shown as a function of $t_\perp/t$ for $U/t=2$,
4, and 8 in Fig.\ \ref{figAFHFgap}.
The transition from a two--band metal to a band insulator in the $U=0$
system occurs at $t_\perp/t=t_{\perp c} = 2.0$.
As $U$ is turned on for $t < t_{\perp c}$ in the AFHF, a gap develops
in both the bonding and antibonding bands, leading to a four--band
insulator.
If $t_\perp$ is then increased, $\Delta$ will go to zero at
approximately the noninteracting ($U=0$) $t_{\perp c}/t=2.0$, as seen
for $U/t=2$ in
Fig.\ \ref{figAFHFgap}, leading to a transition from a four--band to a
two--band insulator.
For large $U$ and small $t_\perp$, $\Delta \approx U/2$, as seen for
$U/t=8$ in Fig.\ \ref{figAFHFgap}, so the
Coulomb splitting of the bands will dominate over $U=0$ band structure
effects.
In this case, the gap will go to zero, i.e. the system will undergo
a transition to a band
insulator only when the band structure
splitting $2 t_\perp$, is of the order of the Coulomb splitting, so that
$t_{\perp c} \approx \Delta \approx U/2$.

\begin{figure}
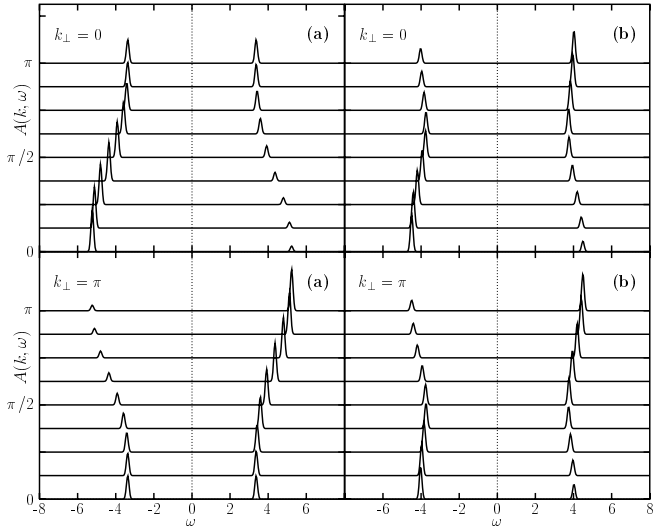

\caption{
        Spectral weight calculated within the AFHF approximation for
	(a) $t_\perp/t=2.0$ and (b) $t_\perp/t=0.5$ with $U/t=8$,
	shown for $k_\perp=0$ and $k_\perp=\pi$.
}
\label{figAFHF}
\end{figure}

The spectral weight within the AFHF approximation is
\begin{equation}
A({\bf k},\omega) = u_{\bf k}^2 \delta (\omega - E_{\bf k})
+ v_{\bf k}^2 \delta (\omega + E_{\bf k}).
\end{equation}
We show
the spectral weight $A({\bf k},\omega)$ from AFHF for the same
parameters as in Fig.\ \ref{figqmc2.0} and Fig.\ \ref{figqmc0.5} in
Fig.\ \ref{figAFHF}(a) for $t_\perp/t=2.0$ and Fig.\ \ref{figAFHF}(b)
for $t_\perp/t=0.5$.
The band splitting $2 \Delta \approx U$ in both cases, as can also be
seen in Fig.\ \ref{figAFHFgap}.
Since both the bonding and antibonding bands are split by this gap,
there are two peaks in $A({\bf k},\omega)$ as a function of $\omega$
for each ${\bf k}$, leading to a four--band insulator for both values
of $t_\perp/t$.
Since the band splitting is set by $U$ rather than $2 t_\perp$, the
spectral weight is
almost evenly distributed between two coherent bands for both
$t_\perp/t=2.0$ and $t_\perp/t=0.5$.
Therefore, for $t_\perp/t=2.0$, as can be seen by comparing
the QMC results in
Fig.\ \ref{figqmc2.0} with Fig.\ \ref{figAFHF}(a), the AFHF spectral
weight distribution for $t_\perp/t=2.0$ is
completely wrong, with too much weight in the upper, $\omega>0$, band
for $k_\perp=0$ and the $\omega < 0$ band for $k_\perp=\pi$.
While the average positions of the AFHF bands, shown as thick solid
lines in Fig.\ \ref{figqmc2.0}(c), are approximately the same as those of
the QMC bands, the band widths are too small by approximately a factor
of two.
For $t_\perp/t=2.0$ and $U/t=8$, then, the AFHF calculation results in a
four--band insulator in which there is long--range antiferromagnetic
order, while the QMC results show that the system is a two--band insulator,
with only very short range antiferromagnetic correlations.
While the AFHF picture does produce a two--band insulator for
$t_\perp/t$ {\raise3pt\hbox{$>$}\llap{\lower3pt\hbox{$\sim $}}} $4.5$,
the physical picture of the transition to this
phase is quite different (see section C).

For $t_\perp/t=0.5$, as seen in Fig.\ \ref{figqmc0.5}
and Fig.\ \ref{figAFHF}(b), AFHF
gives two dispersive bands in the $k_\perp=0$ branch of width of order
$J \sim 4 t^2/U$, similar to the coherent bands seen in the QMC data.
However, in the AFHF, the single particle gap is somewhat too large,
the weight distribution extends too far towards $k=\pi$ and too far
towards $k=0$ for the photoemission and inverse photoemission parts of
the spectrum, respectively, and the broad incoherent
background is not present.
We have also carried out mean--field slave boson
calculations\cite{slaveboson} in order to try to improve on the AFHF
picture.
These calculations give a gap about 15\% smaller than AFHF, and
a bandwidth about 3\% smaller, but do not qualitatively improve on the
AFHF calculations.
The same holds for $t_\perp/t=1.0$, where the AFHF describes approximately
the coherent part of the spectrum and also produces a too large single
particle gap
and a slightly different spectral weight distribution.
Therefore, for $t_\perp/t=0.5$ and for $t_\perp/t=1.0$,
AFHF would give a reasonable description
of the dispersion and general spectral weight distribution of the
coherent spin--wave bands, if the gap were
phenomenologically adjusted to a smaller value, but fails to even
qualitatively describe the spectral weight distribution for
$t_\perp/t=2.0$.

\subsection{\label{RUNGSINGLET} Local Rung Approximation}
\begin{figure}
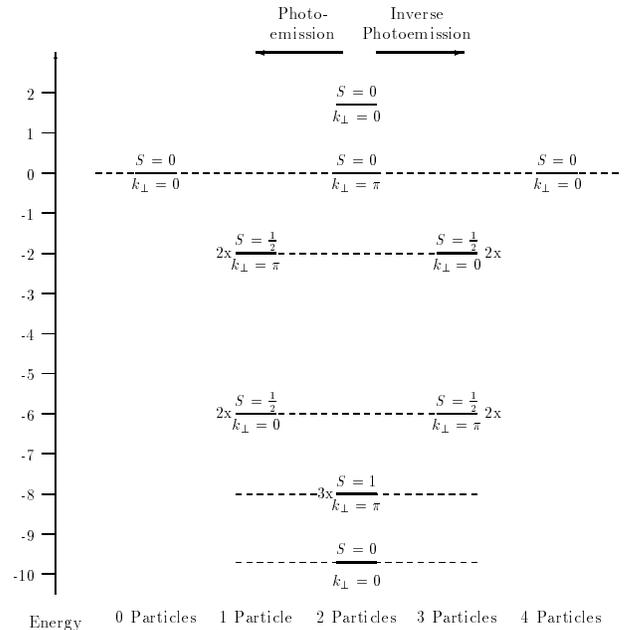

\caption{
	Exact eigenstates for a two-site system representing a
	disconnected rung, with $U=8$ and
	$t_\perp=2.0$.
	The horizontal axis is the number of particles
	and the vertical axis shows the total energy
	$\langle (H-\mu N) \rangle$, where the
	chemical potential $\mu=U/2$ at half-filling.
	Each level is labeled with the
	total spin $S$ and momentum $k_\perp$.
}
\label{figrunglevels}
\end{figure}

For very large $t_\perp$, a better starting point is the
limit of weak interaction between the rungs.
In this limit, we split the Hamiltonian of Eq.\ (\ref{hamiltonian})
into $H=H_0 + H_I$ with
\begin{eqnarray}
H_0&=&-t_\perp \sum_{i,\sigma} (c^\dagger_{i,1\sigma}
               c^{\phantom{\dagger}}_{i,2\sigma} + h.c.)
    + U\sum_{i\lambda} n_{i,\lambda\uparrow} n_{i,\lambda\downarrow},
\nonumber\\
H_I&=&-t\sum_{i,\lambda\sigma} (c^\dagger_{i,\lambda\sigma}
               c^{\phantom{\dagger}}_{i+1,\lambda\sigma} + h.c.),
\end{eqnarray}
where $H_0$ denotes the non--interacting rung limit.
The ground state of $H_0$ is a product of individual rung eigenstates.
Accordingly, we first diagonalize
the Hubbard Hamiltonian on the two sites making up the
rung in order to find the rung eigenstates.
A schematic diagram of the exact eigenstates for a single rung is
shown in Fig.\ \ref{figrunglevels}.
The ground state for the half--filled rung $i$ (two particles per
rung) is a spin singlet state
with $k_\perp=0$ and energy
$E_0/L = E_0^{\text{rung}} = \left( U- \sqrt{U^2+ 16 t_\perp^2} \right)/2$
denoted by $\vert S_i\rangle$.
For large values of
$U$, the energy of this two-site state is given by the exchange
coupling $-J_\perp \sim -4t_\perp^2/U$.
Note that $\vert S_i\rangle$ contains terms which have
double occupied sites as well as the usual singlet construction
$(\vert\uparrow,\downarrow\rangle -\vert\downarrow,\uparrow\rangle)/\sqrt 2$.

We can then form an approximation to the ground state for the
half--filled system by
taking a state which is just a product of the rung states,
\begin{equation}
|\psi_0\rangle = | S_1 \rangle | S_2 \rangle ... | S_L \rangle .
\label{eqsingstate}
\end{equation}
For $t_\perp/t=2.0$, the binding energy of a singlet formed on a rung
is about four times lower than the energy of a singlet between
neighboring sites along a chain, and therefore we
expect $|\psi_0\rangle$ to be a good approximation
to the exact ground state.

\begin{figure}
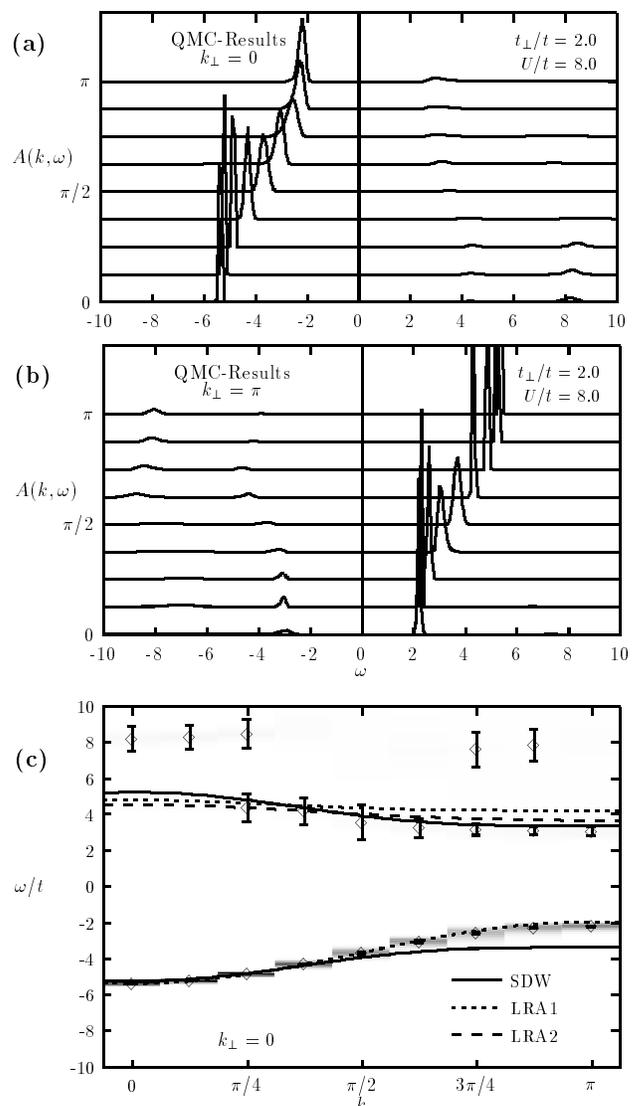

\caption{
        QMC result for the
        spectral density function $A({\bf k},\omega)$ for all possible
	values of ${\bf k}=(k,k_\perp)$ on a $2\times 16$ system with $U/t=8$
	and $t_\perp/t=2.0$ at an inverse temperature of $\beta=10/t$.
	Parts (a) and (b) are three--dimensional plots on the
	$\omega$--$k$ plane for $k_\perp=0$ and $k_\perp=\pi$,
	respectively.
	Part (c) is a density plot for the $k_\perp=0$ branch with
	darker shading corresponding to higher spectral weight, and
	the points with error bars showing the position of the peaks.
	The thick solid line indicates the SDW result within AFHF at
	$T=0$, the
	dashed line is the calculated local rung approximation (LRA)
	dispersion for local excitations on one rung (LRA1) and
	the long--dashed line shows the spectrum obtained
	by local excitations on a four-site cluster (LRA2).
}
\label{figqmc2.0}
\end{figure}

\begin{figure}
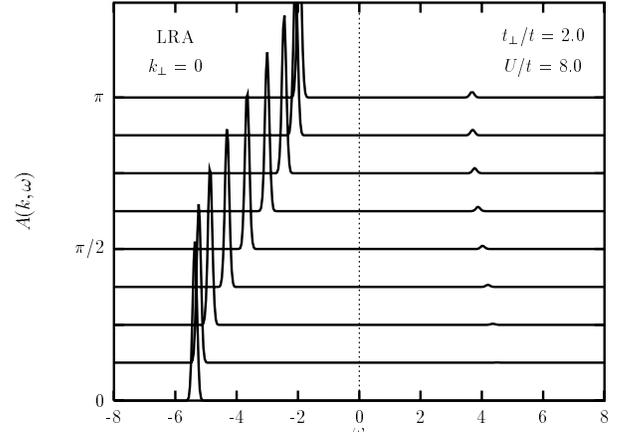

\caption{
	Spectral weight using LRA1 calculation for the $\omega < 0$ and
	the LRA2 calculation for the $\omega > 0$ parts of the
	spectrum for $k_\perp=0$.
	Here $t_\perp/t=2.0$ and $U/t=8$.
}
\label{figLRA}
\end{figure}

In order to obtain the $\omega<0$  spectral weight for the
half--filled system, we need to calculate the matrix element of
$c_{{\bf k}, \sigma}$ given in Eq.\ (\ref{Akomegadef}).
In other words, we need matrix elements between a half--filled
state and a state with one particle removed.
We form the approximate one hole state by replacing one bond singlet
state at rung $\ell$,  $|S_\ell\rangle$, with the lowest energy
one--particle state, the one with bonding ($k_\perp=0$) symmetry,
$|B_{\ell\uparrow}\rangle$
(We remove a spin down electron for definiteness.) and define
\begin{equation}
|\ell\rangle = | S_1 \rangle | S_2 \rangle ...
|B_{\ell\uparrow}\rangle ...| S_L \rangle.
\label{eqlocalsing}
\end{equation}
We delocalize the single--particle state with a plane wave ansatz by
constructing the state\cite{barnes1,bos93}
\begin{equation}
|\psi_1(k)\rangle = L^{-1/2}\sum_{\ell=1}^L e^{ik\ell} |\ell\rangle.
\label{eqdelocalsing}
\end{equation}
Trivially, this state is an exact eigenstate of $H_0$.
We will take this state as the ground state for the system with one
hole with momentum ${\bf k} = (k,0)$.
The spectral weight $A({\bf k},\omega)$ at zero
temperature ($\beta\rightarrow\infty$) is then approximated
by using
only the two states  $|\psi_0\rangle$ and $|\psi_1(k)\rangle$ in
Eq.\ (\ref{Akomegadef}).
For $k_\perp=0$,
the energy dispersion $\omega(k)$ of $A({\bf k},\omega)$ for
$\omega < 0$ is given by
\begin{eqnarray}
\omega(k) &=&\langle\psi_0\vert H \vert\psi_0\rangle -
\langle \psi_1(k)\vert H \vert\psi_1(k) \rangle - \mu \nonumber\\
&=&-\frac{1}{2} \sqrt{U^2+ 16 t_\perp^2} + t_\perp -tA \cos k,
\label{eqLRAwk}
\end{eqnarray}
with
$A=(1+E_2/2t_\perp)^2/(1 + E_2^2/4t_\perp^2)$,
$E_2=( U + \sqrt{U^2+ 16 t_\perp^2})/2$,
and the corresponding spectral weight by
$\vert\langle\psi_1(k)\vert c_{k,\downarrow} \vert\psi_0 \rangle\vert^2$.
The dispersion given by Eq.\ (\ref{eqLRAwk})
for $U/t=8$ and $t_\perp/t=2.0$ is plotted in
Fig.\ \ref{figqmc2.0}(c) and the spectral weight for
$k_\perp=0$ is shown in  Fig.\ \ref{figLRA}.
One can see that the
position of the peak of the $\omega < 0$ LRA band, denoted LRA1
in Fig.\ \ref{figqmc2.0}(c), lays almost exactly on the QMC
data.
However, the position and the dispersion of the $\omega > 0$ band,
which were obtained in
the same way as for $\omega<0$, are not exactly matched by
the LRA1 calculation.
While this band is not very important in the sense that it contains
very little spectral weight, we can understand how to improve
the LRA1 calculation by
considering the states of a four site cluster.

In order to generate the inverse photoemission spectrum, we need to
calculate matrix elements between the half--filled state and a state
with one {\it additional} particle.
{}From Fig.\ \ref{figrunglevels}, we see that a rung state with three
particles and $k_\perp=0$ has approximately twice the
excitation energy of
the $k_\perp=0$ one--particle state which is relevant for the
photoemission part of the spectrum.
(Recall that due to the particle--hole symmetry at half--filling, this
state will map to one in the inverse photoemission part of spectrum for
$k=\pi$; one can see this symmetry in Fig.\ \ref{figrunglevels}.)
Since the relevant three--particle single rung state is high in energy,
configurations involving intrachain effects might also be important.
We include the effect of such configurations by replacing two rung
singlet states by the lowest energy state of five particles on four
sites in the bonding channel.
We can then form a delocalized plane wave state from this state as
in Eq.\ (\ref{eqdelocalsing}).
The results of this calculation are labeled as LRA2 in
Fig.\ \ref{figqmc2.0}(c).
The location and width of the $k_\perp=0$ inverse photoemission band
are more closely fit than by the LRA1 calculation.
In Fig.\ \ref{figLRA} the spectral weight distribution of the LRA1
calculation for the photoemission part of the spectrum ($\omega<0$) and the
distribution of the LRA2 calculation for the inverse photoemission
part ($\omega>0$) is plotted. The result is in good accordance with the
QMC data in Fig.\ \ref{figqmc2.0}(a).

\section{\label{SPINCHARGE} SPIN AND CHARGE DYNAMIC CORRELATION FUNCTIONS}

In order to determine the nature of the low lying excitations, we also
consider the spin and charge susceptibilities
$\chi_{s,c}({\bf q},\omega)$ which, in a Lehmann representation, are defined as
\begin{eqnarray}
\chi_{s,c}({\bf q},\omega)&=&
   {i\over Z}\sum_{l,l^\prime}e^{-\beta E_l}
   (1-e^{-\beta\omega})
   \vert\langle l\vert O_{s,c}({\bf q})\vert l^\prime\rangle\vert^2
\nonumber\\
   &&\phantom{{i\pi\over Z}\sum_{l,l^\prime}}
   \cdot\delta(\omega-(E_{l^\prime}-E_l)),
\label{eqtwocorrdef}
\end{eqnarray}
with $O_{s}({\bf q})=\sum_{\bf p}
(c^\dagger_{{\bf p}+{\bf q},\uparrow}c^{\phantom{\dagger}}_{{\bf p},\uparrow}
-c^\dagger_{{\bf p}+{\bf q},\downarrow}
c^{\phantom{\dagger}}_{{\bf p},\downarrow})$
and
$O_{c}({\bf q})=\sum_{{\bf p},\sigma} c^\dagger_{{\bf p}
+{\bf q},\sigma}c^{\phantom{\dagger}}_{{\bf p},\sigma}$.
We calculate the two--particle dynamic response from the two--particle
imaginary time Green's function, as in Eq.\ (\ref{eqGtaudef}) and use
the Maximum Entropy method described above for the analytical
continuation of $\chi_{s,c}({\bf q},\omega)$ to real frequencies.

\begin{figure}
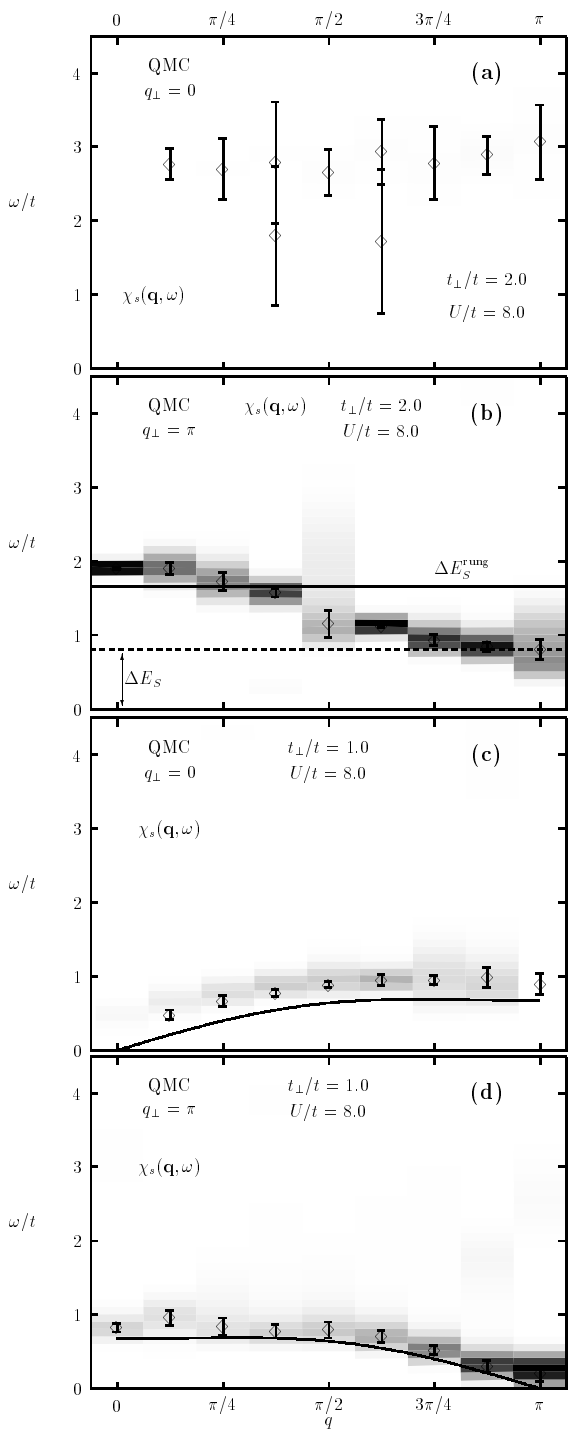

\caption{
	Dynamic spin response function $\chi_s({\bf q},\omega)$
	for (a) $t_\perp/t=2.0$, $q_\perp=0$, (b) $t_\perp/t=2.0$,
	$q_\perp=\pi$, (c) $t_\perp/t=1.0$, $q_\perp=0$, and (d)
	$t_\perp/t=1.0$, $q_\perp=\pi$.
	Again, $U/t=8$ on a $2\times 16$ lattice.
	The weight of sizable structures is represented by the
	strength of shading in the shaded
	regions and the position of the maxima by the points with
	error bars.
	In (b), the solid line indicates the spin gap
	$\Delta E^{\text{rung}}_S$ of the two-site system and the
	dashed line the spin gap
	$\Delta E_S$ for the
	$2\times 16$ ladder at $T=0$, calculated with a projector QMC
	algorithm.
	In (c) and (d), the solid lines are
	$\chi^{+-}_{\rm RPA}({\bf q},{\bf q},\omega)$ calculated using
	the SDW approximation given by Eq.\ (19).
}
\label{figspin}
\end{figure}

In Fig. \ref{figspin} we show the dynamical spin--spin correlation
function $\chi_{s,c}({\bf q},\omega)$ for
$t_\perp/t=2.0$ and $t_\perp/t=1.0$.
For $t_\perp/t=2.0$, the $q_\perp=0$ component [Fig.\ \ref{figspin}(a)]
has a broad, lightly weighted structure centered approximately at
$\omega=3t$, with the spectral weight vanishing for small $q$.
For $q_\perp=\pi$ [Fig.\ \ref{figspin}(b)],
there is a coherent, dispersive band with a width set by
$J \approx 4 t^2/U = 0.5t$, a minimum at $q=\pi$ and a maximum at
$q=0$.
The minimum spin gap is approximately $0.8t$, which agrees well with
the value we obtain using Projector Quantum Monte Carlo, indicated by
the dashed line on the plot.
The spectral weight is most heavily concentrated around $q=\pi$.

The dynamic spin response at half--filling measures the response of
the system to spin triplet excitations.
Since here $t_\perp$ is large, we consider the effect of triplet
excitations on product rung states of the type considered in
Eq.\ (\ref{eqsingstate}), in the LRA calculation.
A triplet excitation will change the total spin of the state from
$S=0$ to $S=1$.
Referring to Fig.\ \ref{figrunglevels}, the only triplet
excited state on a single rung has momentum $\pi$ (corresponding to
odd parity under chain interchange) leading to a change
in momentum $q_\perp=\pi$ for the triplet excitation.
The size of this triplet excitation $\Delta E^{\text{rung}}_{\text{s}}$
is marked on Fig.\ \ref{figspin}(b) by the solid line.
This local triplet excited state can be moved to a near neighbor rung
by a process which is second order in $H_I$.
The local triplet excited states can then be delocalized into a Bloch
wave as in Eq.\ (\ref{eqlocalsing}) and Eq.\ (\ref{eqdelocalsing}).
For large $U$, one then obtains a dispersion relation of the form
$\Delta E^{\text{rung}}_{\text{s}} + J \cos q$,
where $J\sim 4t^2/U$, a form similar to that obtained in
Ref.\ 5 
for the two chain Heisenberg model.
The coherent band in Fig.\ \ref{figspin}(b) does have a minimum at $q=\pi$,
and has approximately this form.

Within the LRA picture, it is also possible to excite a $q_\perp=0$
(even under chain interchange) triplet excitation via a
``two--magnon'' process, as discussed in Ref.\ 5.
This corresponds to making two local triplet excitations on rungs in
the non--interacting rung picture, and would lead to an excitation
energy whose scale, in the Heisenberg limit, is set by
$\omega \sim 2J_\perp$ rather than $\omega \sim J_\perp$.
The overall position of the lightly weighted band seen for $q_\perp=0$
in Fig.\ \ref{figspin}(a) is consistent with this two--magnon band,
although the band is too lightly weighted and our resolution too low
to extract a dispersion relation.

For the physical relevant isotropic case
($t_\perp/t=1.0$), shown in Fig.\ \ref{figspin}(c) and (d),
the spin response looks quite different. (Because the $t_\perp/t=0.5$
results are qualitatively similar to the $t_\perp/t=1.0$ results,
we show only the isotropic ($t_\perp/t=1.0$) case here.)
In Fig.\ \ref{figspin}(c) and (d),
there are dispersive bands in both the $q_\perp=0$ and
$q_\perp=\pi$ branches, with the position of the peak going to a
finite minimum at ${\bf q}=(0,\pi)$ and ${\bf q}=(\pi,0)$, and
appears to vanish as ${\bf q} \rightarrow (\pi,\pi)$.
However, as shown in Fig.\ \ref{figDMRG}, the correlation length
is 4.3 lattice spacings in this regime, and there should be a
spin gap.
Using DMRG calculations on lattices of $2 \times 8$ to $2 \times 32$
sites\cite{nws94}, we estimate the spin gap to be of order $0.12t$.
Due to finite size effects and finite resolution in the analytic
continuation, this gap is too small to resolve in Fig.\ \ref{figspin}(d).
Near ${\bf q}=(0,0)$ [Fig.\ \ref{figspin}(c)], the dispersion of the
peak is hard to discern because there is very little
spectral weight.
This lack of spectral weight at ${\bf q}=(0,0)$ is present in all the
dynamic charge and spin correlations and is due to
a selection rule that comes about because $O_{s,c}({\bf q=0})$
in Eq.\ (\ref{eqtwocorrdef}) commutes with the Hamiltonian, leading
to a vanishing of the matrix element of $O_{s,c}({\bf q})$ as
${\bf q} \rightarrow 0$.
Therefore, the system has relatively long--range spin order in this
regime and, to within the resolution of our calculations, shows
the characteristics of an ordered state with gapless excitations.
Thus it is interesting to compare the QMC spectra to the results of
spin--wave theory calculations.

In order to extract the low--lying spin excitations using spin--wave theory
it is necessary to
consider the RPA transverse spin susceptibility
$\chi^{+-}_{\rm RPA}({\bf k},{\bf k}^\prime,\omega)$.
It is obtained
by applying the random--phase approximation to the response function
$\chi^{+-}_0({\bf k},{\bf k}^\prime,\omega)$, which is calculated directly
in real time using the AFHF ground state $\vert\Omega\rangle$,
calculated in section \ref{AFHFSEC}:
\begin{equation}
\chi^{+-}_0({\bf q},{\bf q}^\prime,t)=
{i\over 4L} \langle\Omega\vert T S^+_{\bf q}(t)S^-_{-{\bf q}^\prime}
(0)\vert\Omega\rangle,
\end{equation}
with $S^\pm=S_x\pm iS_y$.
Due to the broken spin rotational symmetry of $\vert\Omega\rangle$,
$\chi^{+-}_{\rm RPA}({\bf q},{\bf q}^\prime,\omega)$ contains a gapless mode,
as predicted by the Goldstone theorem.
Following the procedure of Schrieffer et al.\cite{swz89}, one obtains:
\begin{eqnarray}
\lefteqn{
\chi^{+-}_{\rm RPA}({\bf q},{\bf q}^\prime,\omega)=}\ \ \ \ \ \nonumber\\
&&\sum_{\bf p} \chi_0^{+-}({\bf q},{\bf p},\omega)
[1-U\chi_0^{+-}({\bf p},{\bf q}^\prime,\omega)]^{-1},
\end{eqnarray}
where $[1-U\chi_0^{+-}({\bf p},{\bf q}^\prime,\omega)]^{-1}$ is a
matrix inverse of a $2\times 2$ matrix in ${\bf q}$-space and
\begin{eqnarray}
\chi^{+-}_0({\bf q},{\bf q}^\prime,\omega)&=&
\delta({\bf q}-{\bf q}^\prime)\chi_0^{+-}({\bf q},\omega)\nonumber\\
&&+ \delta({\bf q}-{\bf q}^\prime+{\bf Q})\chi_Q^{+-}({\bf q},\omega),
\end{eqnarray}
with
$\chi^{+-}_0({\bf q},\omega)$ and $\chi^{+-}_Q({\bf q},\omega)$ given
by the usual ``bubble'' diagrams and printed in detail in
Ref.\  16 
[up to a misprint of a factor of 2 in $\chi^{+-}_Q({\bf q},\omega)$].
The dispersion of $\chi^{+-}_{\rm RPA}({\bf q},{\bf q},\omega)$ is indicated
as a solid line in Fig.\ \ref{figspin}(c) and \ref{figspin}(d).
The RPA spin--wave dispersion goes to zero at ${\bf q}=(0,0)$ and
${\bf q}=(\pi,\pi)$, and goes to a finite minimum at ${\bf q}=(0,\pi)$
and ${\bf q}=(\pi,0)$, consistent with the QMC data.
The spin--wave velocity, the width of the bands, and the gaps at
${\bf q}=(0,\pi)$ and ${\bf q}=(\pi,0)$ are also in
reasonable agreement with the QMC data.

\begin{figure}
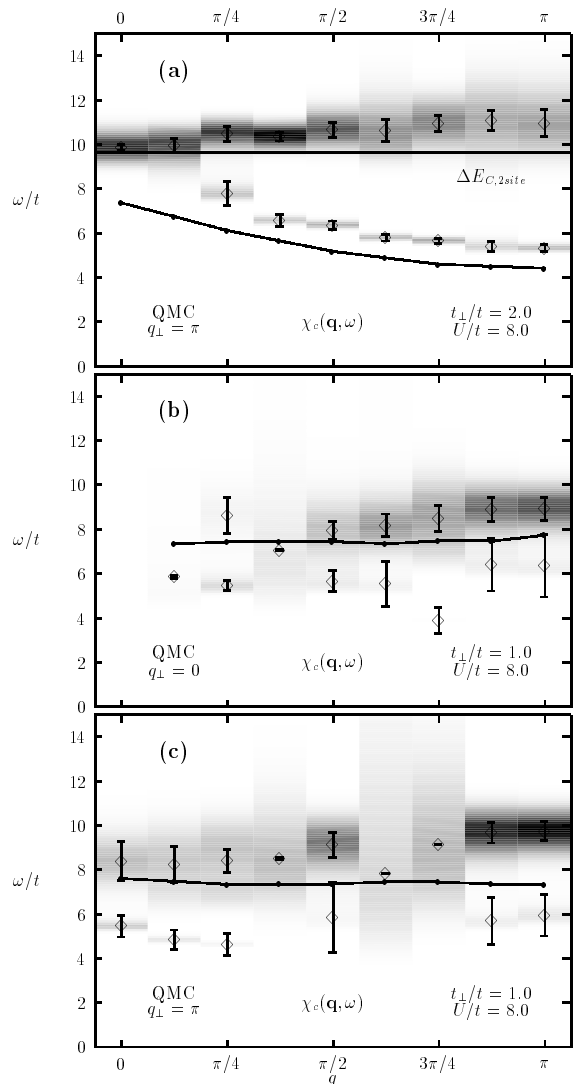

\caption{
	The dynamic charge
	response function $\chi_c({\bf q},\omega)$ for the same
	parameters as in Fig.\ 7.
	In (a), $t_\perp/t = 2.0$ and $q_\perp=\pi$, the solid line
	represents the charge excitation energy
	on a single rung $\Delta E^{\text{rung}}_C$, and the line with
	solid dots represents the minimum excitation energy
	constructed from the single particle spectral weight in
	Fig.\ 1.
	In (b) and (c), $t_\perp/t = 1.0$, and $q_\perp=0$ and
	$\pi$, respectively.
	The lines with solid dots represent the
	minimum excitation energy of the RPA charge susceptibility
	$\chi^{00}_{\text{RPA}}({\bf q},{\bf q},\omega)$
	calculated with the method of  Ref.\ 16.
}
\label{figcharge}
\end{figure}

The QMC results for the dynamic charge correlation function are shown in
Fig.\ \ref{figcharge} for $t_\perp/t=2.0$ and $t_\perp/t=1.0$.
For $t_\perp/t=2.0$, almost all of the spectral weight is in the
$q_\perp=\pi$ component, so we do not show the $q_\perp=0$ component.
There exist two important features: one is a heavily weighted,
relatively flat band at $\omega \sim 10t$ with heaviest weight near $q=0$.
This band becomes somewhat incoherent as $q$ increases.
The second is a flat, dispersive, less heavily weighted band
with a minimum of order $5t$ near $q=\pi$.
The spectral weight in this band extends from about $q=\pi/4$ to $q=\pi$,
and the size of the charge gap is set by this lightly weighted lower band.

In the large $t_\perp$ limit,
one can understand the structure of the charge response from the LRA
picture described in section \ref{RUNGSINGLET}.
In the single rung picture, a charge excitation will occur through a
transition to an excited state conserving the number of particles on
the rung and the total spin; in
other words, the important transition will be in the middle column of
Fig.\ \ref{figrunglevels}, from the low-lying $S=0$ state to the
higher $S=0$ states.
There are two possible charge excited states on the rung, one with
momentum $k_\perp=0$ (even parity) and one with momentum
$k_\perp=\pi$ (odd parity).
In a single rung picture, an optical transition from the $k_\perp=0$
ground state to a $k_\perp=0$ excited state is forbidden
because the $q_\perp=0$ density operator $O_c(0)$ commutes with
the Hamiltonian.
This selection rule forbidding a $q_\perp=0$ optical transition
remains present when the rung charge excited states are constructed as
in Eq.\ (\ref{eqlocalsing}) and delocalized in a
state like that in Eq.\ (\ref{eqdelocalsing}).
This is why there is almost no spectral weight in the $q_\perp=0$
portion of the charge response for $t_\perp/t=2.0$.
Of course, the system is not exactly in a LRA state, so there will
be some higher order processes that will introduce a very small amount
of spectral weight into the $q_\perp=0$ branch.
The energy of the $q_\perp=\pi$ single rung transition, indicated by a
solid line on the Fig.\ \ref{figcharge}(a), gives an excitation energy
that agrees well with energy of the heavily weighted region at $q=0$.

In order to qualitatively understand the origin of the dispersion of
the charge
response, one can consider the possible particle--hole excitations
within the one--particle band structure given by
$A({\bf k},\omega)$ in Fig.\ \ref{figqmc2.0}.
There will be significant amplitude in the two--particle charge
response when there is significant amplitude for a transition at a
particular ${\bf q}=(q,q_\perp)$ and $\omega$ for a particle--hole
excitation built up from the one--particle spectral weight.
For example, to understand the heavily weighted amplitude at
${\bf q} = (0,\pi)$, one has to integrate over all transitions from
the photoemission band in Fig.\ \ref{figqmc2.0}(a) to the inverse
photoemission band in Fig.\ \ref{figqmc2.0}(b) which transfer this
momentum.
Since the single--particle bands are sharp and parallel, one should
obtain a single well--defined peak for ${\bf q} = (0,\pi)$, which we
see in Fig.\ \ref{figcharge}(a) by considering excitations between
the $\omega<0$, $k_\perp=0$ band, and the $\omega>0$, $k_\perp=\pi$
band.
In addition, the transition at ${\bf q} = (0,\pi)$ has odd parity
between the chains, and is thus allowed by the selection rules for the
density operator as $q\rightarrow 0$.
As the parallel component $q$ is increased, one can see that there
will be a continuum of excitation energies which gets wider as $q$
increases.
The minimum excitation energy (position of the lower band) from this
construction is shown in Fig.\ \ref{figcharge}(a) by the line with
solid dots.
However, the excitation energy obtained
is consistently smaller than the energy of the lowest heavily
weighted band from QMC.
Using the single particle bands to construct the two--particle
excitations is equivalent to calculating the charge response using the
lowest order ``bubble'' diagram, but with exact single--particle
propagators, neglecting all particle--hole interactions.
The particle--hole interactions on the rung, which are included in the
rung eigenenergies, thus raise the charge
excitation energy by a substantial amount\cite{oldhanke}.

For $t_\perp/t=1.0$, the charge response looks quite different.
As shown in Fig.\ \ref{figcharge}(b) and (c),
there is substantial spectral weight for both
$q_\perp=0$ and $q_\perp=\pi$.
For ${\bf q} \rightarrow (0,0)$, the density operator has even parity,
causing the matrix element and thus the spectral weight to
vanish, whereas
at ${\bf q} \rightarrow (0,\pi)$, the density operator has odd parity
so that optical transitions are allowed and there is spectral weight.
In both channels, the size of the charge gap is approximately $4t$,
and there is a broad structure of width $\sim 7t$.
For $q_\perp=0$, most of the spectral weight occurs as a dispersive peak
whose energy increases with
increasing $q$, whereas for $q_\perp=\pi$  two peaks seem
to contribute to the spectral weight distribution.
The peaks are not well-defined enough over a range of $q$ to extract a
dispersion, but the upper peak is heavily weighted near $q=\pi$, at
$\omega \approx 9t$.

We can qualitatively understand the broad incoherent structure of the
charge response by considering particle--hole excitation in the single
particle $A({\bf k},\omega)$ in Fig.\ \ref{figqmc1.0}.
There are four dispersive bands and a broad background, so there
should be weight in both the $q_\perp=0$ and $q_\perp=\pi$ branches of
the charge response, and broad structure above a minimum excitation
energy, which we see in Fig.\ \ref{figcharge}(c) and (d).
 From the single particle bands in Fig.\ \ref{figqmc0.5}, one can
estimate the minimum particle--hole excitation energy to be
$\sim 4t$.
In Fig.\ \ref{figcharge}(c) and (d), the spectral weight near this
minimum excitation energy is suppressed due to the particle--hole vertex.
We have also carried out a calculation of charge response
$\chi^{00}_{\text{RPA}}({\bf q},{\bf q}^\prime,\omega)$ within the
antiferromagnetic RPA approximation described above, using the SDW
dispersion in Eq.\ (\ref{eqAFHFdisp}), and also find a
relatively broad structure above the charge gap for both
$q_\perp=0$ and $q_\perp=\pi$.
We plot the minimum excitation energy of
$\chi^{00}_{\text{RPA}}({\bf q},{\bf q}^\prime,\omega)$ in
Fig.\ \ref{figcharge}(b) and (c) as lines with solid dots.
This line is located in the middle of the broad band in both plots.

The spin and charge response functions for $t_\perp/t=0.5$ show the same
general
features as in the isotropic ($t_\perp/t=1.0$) case and are therefore not shown
here. The spin susceptibility $\chi_s({\bf q},\omega)$ is also
qualitatively identical to the RPA result
$\chi^{+-}_{\rm RPA}({\bf q},{\bf q},\omega)$ with a smaller spin velocity
than in the $t_\perp/t=1.0$ case. The charge susceptibility $\chi_c({\bf
q},\omega)$ shows a clear dispersive band centered around the low--lying
RPA excitations in the $q_\perp=0$ channel, whose energy increases with
increasing $q$. In the $q_\perp=\pi$ channel, there is a broad structure
of width $\sim 8t$ with again two peaks in the spectral weight distribution.

\section{\label{CON} CONCLUSION}

In summary, the single and two--particle dynamical properties of the
two chain Hubbard model at half--filling can be understood by starting
from two limits: the limit of non--interacting rungs treated exactly,
which gives a good starting point for the large $t_\perp$ case for
which the spin--spin correlation length along the chains is less than
a lattice spacing, and
the limit of an antiferromagnetically ordered state, which gives a
good starting point for the small $t_\perp$ case when the spin--spin
correlation length is large.
The dynamical properties in the two regimes look quite different.
For the large $t_\perp$ regime, the remnants of the level transitions
of the two site system representing a single rung, suitably broadened
into bands, can explain the major features of
the single--particle, spin and charge responses.
In the small $t_\perp$ case, calculations based on an
antiferromagnetically ordered starting point, such as
antiferromagnetic Hartree--Fock theory and spin--wave theory give a
good qualitative picture of the coherent spin--wave part of the single
particle spectral weight and of the two--particle spin dynamic
correlation function.
In addition, there is a broad incoherent band in the single particle
spectral weight similar to that found in recent numerical work on the
1D and 2D systems.

We also have shown the single particle spectral weight and the spin and charge
response functions for the physical relevant, isotropic case
($t_\perp/t=1.0$). The results are qualitatively similar to these in the small
$t_\perp$ region and can therefore be understood from an antiferromagnetic
Hartree--Fock or spin--wave theory picture. The real ladder compounds
have approximately the same coupling strength parallel to and perpendicular
to the chains, and it will therefore be interesting to
compare these isotropic $t_\perp/t=1.0$ results with future
experiments.

\section*{ACKNOWLEDGMENTS}

We would like to thank R.\ Preuss, A.\ Muramatsu, and W.\ Ziegler for
helpful discussions.
H.E., W.H., and R.M.N.\ are grateful to the Bavarian "FORSUPRA" program
on high $T_c$ research and the DFG under Grant No.\ Ha 1537/12-1 for financial
support.
D.P. acknowledges support from the EEC Human Capital and Mobility
program under grant CHRX-CT93-0332, and D.J.S. from the NSF under
Grant No.\ DMR 92-2507.
The calculations were performed on Cray YMP's at the HLRZ in
J\"ulich and at the LRZ M\"unchen.

\begin{figure}
\caption{
	The single particle spectral weight $A({\bf k},\omega)$ as in
	Fig.\ 1, 
	with the same parameters except with
	$t_\perp/t=0.5$.
	In (c), the solid line is the SDW result
	calculated within AFHF.
}
\label{figqmc0.5}
\end{figure}

\newpage
\begin{thebibliography}{999}
\bibitem{takano} Z. Hiroi, M. Azuma, M. Takano, and Y.\ Bando, J. Solid
    State Chem., {\bf 95}, 230 (1991).
\bibitem{johnston} D.C. Johnston {\it et al.}, \prb {\bf 35}, 219 (1987).
\bibitem{rice} T.M. Rice, S. Gopalan, and M. Sigrist, Europhys. Lett.
    {\bf 23}, 445 (1993).
\bibitem{hit95} Z.\ Hiroi, and M.\ Takano, Nature {\bf 377}, 41 (1995).
\bibitem{dagotto} E.\ Dagotto, J.\ Riera, and D.J.\ Scalapino, \prb
    {\bf 45}, 5744 (1992).
\bibitem{barnes1} T.\ Barnes et al., \prb {\bf 47}, 3196 (1993);
    T.\ Barnes and J.\ Riera, \prb {\bf 50}, 6817 (1994);
    R.S.\ Eccleston, T.\ Barnes, J.\ Brody, and J.W.\
    Johnson, \prl {\bf 73}, 2626 (1994).
\bibitem{whiteprl} S.R.~White, R.M. Noack, and D.J. Scalapino,
    \prl {\bf 73}, 886 (1994).
\bibitem{nws94}
    R.M.\ Noack, S.R.\ White, and D.J.\ Scalapino,
    Phys.\ Rev.\ Lett.\ {\bf 73}, 882 (1994); Europhys.\ Lett. {\bf 30}
    (3), 163 (1995).
\bibitem{ttr94}
    H.\ Tsunetsugu, M.\ Troyer, and T.M.\ Rice,
    \prb {\bf 49}, 16078 (1994).
\bibitem{maxent}
    R.N.\ Silver, D.S.\ Sivia, and J.E.\ Gubernatis,
    Phys.\ Rev.\ B {\bf 41}, 2380 (1990), and references therein.
\bibitem{baf95}
    L.\ Balents, and M.P.A.\ Fisher, to be published, cond-mat/9504082.
\bibitem{preuss1d}
    R.\ Preuss, A.\ Muramatsu, W.\ von der Linden, P.\ Dieterich,
    F.F.\ Assaad, and W.\ Hanke, Phys.\ Rev.\ Lett.\ {\bf 73},
    732 (1994).
\bibitem{newpreuss1d}
    R.\ Preuss, A.\ Muramatsu, W.\ von der Linden, and W.\ Hanke,
    unpublished.
\bibitem{preuss2d} R.\ Preuss, W.\ Hanke, and
    W.\ von der Linden, to be published, cond-mat/9412089.
\bibitem{dagotto2d} A.\ Moreo, S.\ Haas, A.\ Sandvik, and E.\ Dagotto,
    to be published, cond-mat/9412107.
\bibitem{preussmaxent}
    W.\ von der Linden, R.\ Preuss and W.\ Hanke, to be published,
    cond-mat/9503098.
\bibitem{whi91}
    S.R.\ White, Phys.\ Rev.\ B {\bf 44}, 4670 (1991).
\bibitem{swz89}
    J.R.\ Schrieffer, X.G.\ Wen, and S.C.\ Zhang, Phys.\ Rev.\ B {\bf 39},
    11663 (1989).
\bibitem{kas90}
    A.P.\ Kampf and J.R.\ Schrieffer, Phys.\ Rev.\ B {\bf 42},
    7967 (1990).
\bibitem{slaveboson} L.\ Lilly, A.\ Muramatsu, and W.\ Hanke, \prl
    {\bf 65}, 1379 (1990), and references therein.
\bibitem{bos93}
    I.\ Bose and S.\ Gayan, Phys.\ Rev.\ B {\bf 48}, 10653 (1993).
\bibitem{oldhanke} W.\ Hanke, and L.J.\ Sham, \prl {\bf 43}, 387
    (1979); \prb {\bf 21}, 4656 (1980).
\end{thebibliography}
\end{document}